# WAVE MECHANICS AND THE FIFTH DIMENSION

Paul S. Wesson[1] and James M. Overduin[2,3]

[1]Department of Physics and Astronomy, University of Waterloo, Waterloo, ON, N2L 3G1, Canada.

[2]Department of Physics, Astronomy and Geosciences, Towson University, Towson, MD, 21252, U.S.A.

[3]Department of Physics and Astronomy, Johns Hopkins University, 3400 N. Charles St., Baltimore, MD, 21218, U.S.A.

Abstract: Replacing 4D Minkowski space by 5D canonical space leads to a clearer derivation of the main features of wave mechanics, including the wave function and the velocity of de Broglie waves. Recent tests of wave-particle duality could be adapted to investigate whether de Broglie waves are basically 4D or 5D in nature.

# WAVE MECHANICS AND THE FIFTH DIMENSION

At first sight, it may appear unlikely that higher dimensions can lead to significant effects on local low-energy physics. However, a little thought shows that this can indeed be the case, provided the higher-dimensional manifold is not Minkowski. In this regard, it was recently shown that an exact solution for the so-called five-dimensional canonical metric displays behaviour which mimics that of de Broglie waves [1]. The present account is more general, and will show how the apparently baffling properties of 4D wave mechanics follow naturally from 5D canonical space.

Wave mechanics is currently undergoing a modest revival in the laboratory, because modern technology enables us to carry out new versions of the double-slit experiment and improve our knowledge of wave-particle duality [2]. However, while it is indisputable that particles sometimes act as waves, our theoretical understanding of this phenomenon remains poor. In particular, we do not properly understand the origin and nature of de Broglie waves, or why they apparently travel at velocities which exceed the speed of light. This is curious, given that in recent years there has been tremendous advancement in studies of field theories like general relativity and in the quantum field theory of elementary particles. It is now widely believed that the best route to the eventual unification of these subjects lies in higher dimensions, such as space-time-matter theory and membrane theory [3]. Both of these theories can be formulated in terms of five dimensions, and are in agreement with observations. It is unfortunate that wave mechanics, flanked so to speak by improved theories for physics on larger and smaller scales, should have been relatively neglected. In what follows, it is hoped to remedy this situation.

We will adopt the space-time-matter approach, for which Campbell's embedding theorem ensures the recovery of 4D matter from the fifth dimension, and make use mainly of the canonical metric [4]. The literature on the latter is large, and among other things shows that the timelike paths of massive particles in 4D correspond to *null* paths in 5D [5]. Making use of this and other results, it is hoped to show that the major properties of wave mechanics follow much more naturally from 5D than they do from 4D.

2. The Origin and Nature of de Broglie Waves

In this section, we will take the 5D canonical metric with null interval and from it derive 4D properties which will be found to be the same as those of conventional wave mechanics. This applies in particular to the de Broglie wave, whose velocity is superluminal. This property plus the need for a supporting vacuum clearly demarks de Broglie waves from conventional gravitational waves, which are limited by the speed of light and exist in truly empty space. We will not be concerned with field equations, because it is already known that the 5D (pure) canonical metric corresponds to 4D metrics of general relativity with a finite cosmological constant [1, 4]. In other words, our results will derive ultimately from the nature of the manifold. In this regard, it is important to recall that the 5D canonical metric ($C_5$) is in general not equivalent to the 5D Minkowski metric ($M_5$). The 5D manifold has structure, and so also does the 4D one embedded in it. The paths of test particles and waves in the 4D metric are determined by the null-geodesic condition in the 5D metric. In short, the goal is to derive wave mechanics from the fifth dimension.

The notation is standard. The 5D interval $dS^2$ contains the 4D one $ds^2$, where the 4D proper time $s$ is used as parameter to make contact with conventional physics. The 4D metric tensor $g_{\alpha\beta}(x^\gamma)$ is restricted to be a function only of the spacetime coordinates, but the 4D part of the 5D metric is quadratically factorized in terms of the extra coordinate $x^4 = l$. This defines the typical $C_5$ form, in which 4D physics occurs on the hypersurfaces $l$ = constants. (The manifold resembles a pseudosphere with the 'radius' measured by $l$ and the distance around the 'circumference' measured by $s$.) Dynamical properties of $C_5$ have been much studied [4 – 7]. Typically, when $g_{44}$ = +1, the motion is oscillatory, confined to an $l$ - hypersurface by a negative value of the cosmological constant. This behaviour figures in a recent model which shows a kind of 4D quantization [7]. However, we will not be concerned with this in the present account, though henceforth we will adopt the choice $g_{44}$ = +1 for the potential corresponding to the scalar field of 5D relativity. (This is admissible, since the extra coordinate does not have the physical nature of a time.) The physical constants $c$, $G$ and $h$ will usually be left explicit to aid understanding.

The canonical metric has line element

$$dS^2 = (l/L)^2 ds^2 + dl^2$$

$$ds^2 = g_{\alpha\beta}(x^\gamma) dx^\alpha dx^\beta \quad . \tag{1}$$

Here $L$ is a constant length, which in gravitational problems is related to the cosmological constant [4], but will be given an alternative interpretation below. 5D null-paths with $dS^2 = 0$ in (1) respect 4D causality with $ds^2 \geq 0$, but the fifth dimension has

$$l = l_* \exp(\pm is/L) \quad . \tag{2}$$

This describes a wave oscillating around $l = 0$ with amplitude $l_*$ and wavelength $L$. [The locus of the wave can be shifted to $l = l_0$ if $l$ is changed to $(l - l_0)$, and the sign choice in (2) reflects the reversibility of the motion in the extra dimension.] The function (2) is formally the same as the conventional wave function, if $L = h/mc$ is identified with the Compton wavelength of the associated particle of mass $m$. It should be recalled that the conventional wave function is commonly constructed by taking the complex representation of the action $I = \int mc\,ds$ via $\psi \equiv \exp(\pm iI/h)$. Here

$$I = \int mc\,ds = \int p_\alpha dx^\alpha \quad , \tag{3}$$

where $p^\alpha \equiv mu^\alpha$ is the 4-momentum and $u^\alpha \equiv dx^\alpha/ds$ is the 4-velocity. The momenta can then be obtained from the wave function via $p_\alpha = (\pm h/i\psi)(\partial\psi/\partial x^\alpha)$. The wave function itself obeys the Klein-Gordon equation

$$\Box^2 \psi + (mc/h)^2 \psi = 0 \quad , \tag{4}$$

where $\Box^2 \psi \equiv g^{\alpha\beta}(\partial\psi/\partial x^\alpha)_{;\beta}$ involves the covariant derivative if the spacetime is curved. The Klein-Gordon equation is actually the operator-analog of the normalization condition $E^2 - p^2c^2 - m^2c^4 = 0$ for the energy, momenta and mass of a particle. This, however, is merely the result of appropriate definitions ($E \sim u^0$ etc.), combined with $m^2$ multiplied onto the normalization condition $u^\alpha u_\alpha = 1$ for the 4-velocities in general relativity. And the Klein-Gordon equation (4) may be shown to be equivalent to the extra component of the 5D geodesic equation for the canonical metric (1), or $\delta\left[\int dS\right] = 0$ around the null-path [7]. In fact, we realize that all of the algebraic machinery of conventional 4D wave mechanics is equivalent to the 5D canonical metric.

It is remarkable, in retrospect, that workers like de Broglie in the 1920s were able to piece together a formalism which however cumbersome was able to account for the results of experiments like that of the double slit. Nevertheless, many people remained dissatisfied with wave mechanics, not least because some of its implications appeared to be contrary to special relativity. Prime among these was de Broglie's inference that the velocity of a particle in ordinary space $u$ and the phase velocity of its associated wave $w$ must be related by

$$uw = c^2 \quad . \tag{5}$$

Obviously, for $u < c$ we have $w > c$. Various attempts were made to explain this puzzling result, notably by identifying $u$ with the group velocity of the wave in a strange medium with a finite refractive index, and in terms of hypothetical tachyons with superluminal speeds that populated an otherwise undetectable vacuum [8]. Attempts to show that (5) is consistent with the Lorentz transformations continue to modern times [9]. But while re-cent experiments confirm traditional results [2], it is clear that what is needed is a better theoretical framework. We saw above that the 5D canonical metric (1) with null paths yields an expression (2) equivalent to the old wave function, so below we ask about the associated physics.

The phase velocity of the de Broglie wave turns out to be relatively easy to calculate. Suppose the wave accompanies a particle with rest mass $m$ and momenta components $p^\alpha = mu^\alpha$. The frequency of the wave $f$ is related to $m$ by Planck's law, so

$$hf = mc^2 \quad , \quad f = \frac{mc^2}{h} = \frac{c}{L} \quad . \tag{6}$$

Here $L$ is the constant in the metric (1), which is included there as defining the size of the potential well for 4D spacetime, or can alternately be appended to the extra part of the metric defining the 'size' of the fifth dimension. In either case, $f$ is a scalar quantity. By contrast, the wavelengths $\lambda^\alpha$ define a 4-vector, whose components are by (2) and (3) inversely proportional to the momenta $p^\alpha$. Let us concentrate on the $x$-axis, and write $\lambda_x = h/p^x$. The wave velocity $w_x$ along the $x$-axis is then

$$w_x = f\lambda_x = c^2/u_x \quad , \tag{7}$$

where (6) has been used. That is $u_x w_x = c^2$, agreeing with de Broglie's relation (5). We realize that, according to the present interpretation, the phase velocity of the wave is different along different directions of ordinary space.

This raises questions about the nature of the medium through which the wave propagates. To investigate, we remember that the $C_5$ metric (1) in combination with Einstein's field equations shows that there is a cosmological constant $\Lambda$ present [4], which in gravitational problems determines the pressure and density of the vacuum thus:

$$\Lambda = -\frac{3}{L^2} \quad , \quad p_v = -\rho_v = -\Lambda c^2 / 8\pi G \quad . \tag{8}$$

The equation of state here is typical of the classical vacuum, and should be preserved even when the problem concerned is not gravitational but quantum in nature. However, in that case we expect that the defining parameter should not be $G$ but rather h. The relevant expression for the magnitude of the vacuum density, up to a dimensionless factor, is then $|\rho_v| = h\Lambda^2 / 8\pi c$. Recalling that the metric (1) has $\Lambda = -3/L^2$ and the wave function (2) has $L = h/mc$, the vacuum density can be expressed thus:

$$|\rho_v| = \frac{h\Lambda^2}{8\pi c} = \left(\frac{9}{8\pi}\right)\frac{c^3 m^4}{h^3} = \frac{3}{2}\frac{m}{(4\pi\lambda_C^3/3)} \quad . \tag{9}$$

Here $\lambda_C = h/mc$ is the Compton wavelength. The implication of this expression is clear: the effective density experienced by the wave is the mass of the particle divided by the volume of its Compton sphere.

Logical as this result is, there remain other things about de Broglie waves which need clarifying. For example, are the waves longitudinal (like sound waves) or transverse (like light waves)? To investigate this, let us recall a result about the equations of motion for $C_5$ spaces [6, 7]. The 5D geodesic equation provides in general 4 relations for the motion in spacetime and an extra relation for the motion in $x^4 = l$. The latter relation, as mentioned above, is in general equivalent to the Klein-Gordon equation (4); and for the special case where the metric has the simple form (1) and the path is 5D-null, is also equivalent to $l = l(s)$ as given by (2). The 4 relations for motion in spacetime in general take the form of 4D geodesic motion as found in Einstein's theory plus a perturbation due to the extra dimension. This new force (per unit mass) has been identified in both spacetime-matter theory and membrane theory, whose mathematical structures are similar [10]. It acts parallel to the regular 4-velocity, and is given in general by

$$P^\mu = \left( -\frac{1}{2} \frac{\partial \gamma_{\alpha\beta}}{\partial l} u^\alpha u^\beta \right) \frac{dl}{ds} u^\mu \quad . \tag{10}$$

Here, the term in parentheses represents the coupling between 4D and the extra dimension, and $\gamma_{\alpha\beta} = \gamma_{\alpha\beta}(x^\gamma, l)$ is the 4D metric tensor in unfactorized form. Since the force (per unit mass) is proportional to the relative velocity between the two frames, it is inertial in the Einstein sense. However, $P^\mu$ is gauge-dependent, because it depends on the form of the metric tensor $\gamma_{\alpha\beta}(x^\gamma, l)$ which in turn depends on the choice of coordinates. This is the source of some confusion in the 5D literature, akin to the argument about using the Einstein frame or the Jordan frame in old 4D scalar-tensor theory. In the present situation, there are two choices of coordinates that are relevant: one which decouples the 4D part of the metric form $x^4 = l$, giving $P^\mu = 0$; and the other which keeps the coupling, giving $P^\mu \neq 0$ and resulting in significant dynamics. (A similar situation occurs in standard cosmology, where comoving coordinates means that the galaxies appear to be at rest, while non-comoving coordinates results in the galaxies moving according to Hubble's law.)  Here we wish to examine the finite effects of the acceleration (10), so in that relation we need to put by (1) $\gamma_{\alpha\beta} = (l/L)^2 g_{\alpha\beta}(x^\gamma)$.  This gives a perturbing force

$$P^\mu = -\frac{1}{l}\frac{dl}{ds}u^\mu \quad . \tag{11}$$

The negative sign here reflects a restoring force towards the local centre of motion.  The motion may be obtained by using (2) to substitute for $(1/l)(dl/ds)$, equating $P^\mu$ to the local acceleration $du^\mu/ds$, and integrating.  We omit the details, because the result is just as expected, namely simple harmonic motion with the same wavelength $L = h/mc$ as before.  This motion, it should be noted, represents the perturbation due to the extra dimension; and in a practical situation such as the double-slit experiment, there will in general be another, collimated component of the motion due to the setup of the apparatus.  The acceleration (11), despite its dependency on $x^4 = l$, acts *in* the surface of spacetime, i.e. parallel to *s*. As far as the de Broglie wave is concerned, we conclude that it is of longitudinal type.

3.  <u>Discussion</u>

Wave mechanics is a quaint subject, based on principles which were taken directly from a few experiments performed long ago, and with little of the sophistication that has attended the development of other disciplines such as general relativity and quan-tum field theory. However, recent experiments have confirmed and extended the inferences drawn before; and awkward as they may be, modern physics has to accept that de Broglie waves are in some sense 'real'. While the search proceeds for a theory which unifies gravity with the interactions of particles, it is mainly along routes involving group theory and extra dimensions, and in such a framework wave-particle duality is somewhat anomalous. The present work is motivated by the wish to develop a more logical account of wave mechanics which fits into the scheme of higher-dimensional field theory.

One extra dimension, added to the four of Einstein's theory, is remarkably successful in explaining the nature of matter and particles, under the headings of space-time-matter theory and membrane theory [3, 10]. Notably, the 5D canonical metric embeds 4D spacetime with a kind of spherical symmetry in the higher dimension. This embedding extends to 5D all solutions of Einstein's equations which are empty of ordinary matter but have a finite cosmological constant [4]. This includes the de Sitter solution, which was recently shown via a coordinate transformation to give a good account of de Broglie waves [1]. The present work is more general in scope, and asks if the canonical metric can explain the generic properties of matter waves.

The answer to this is positive. The metric (1), with a null-path, shows that a particle is equivalent to a wave which moves in spacetime in a manner (2) identical to that described by the conventional wave function. A puzzling consequence of conventional wave mechanics is the existence of a superluminal phase velocity (5); but this may be shown by (6) and (7) to follow from the frequency and wavelength of the 5D wave. The medium through which de Broglie waves propagate cannot be observed directly, but it is like the Einstein vacuum. The latter (8), however, can be replaced by another medium whose density (9) is basically given by the particle mass smeared over the Compton wavelength. The nature of a de Broglie wave/particle is related to the extra force (per unit mass) typical of 5D metrics (10). This acts in the surface of spacetime and for our metric (1) takes a particularly simple form (11) which implies that the wave is of longitudinal type.

The preceding discussion shows that the main phenomenological properties of wave mechanics can be understood on the basis of a 5D metric of canonical type. There are, of course, other things to be investigated using modern experimental techniques. The major things are: (a) confirmation that the waves are longitudinal, with observations of the microscopic simple-harmonic motions of the particles; (b) detection of the medium through which the waves propagate, with measurements of its density; (c) investigations of the relation between the ordinary speed of the particle and the phase velocity of its associated wave, to see if indeed something is moving faster than light.

Acknowledgements

Thanks for comments go to members of the Space-Time-Matter group (5Dstm.org).Acknowledgements

Thanks for comments go to members of the Space-Time-Matter group (5Dstm.org).

References

[1]  P.S. Wesson, arXiv: 1301.0333 (2013). See also: arXiv: 1205.4452 (2012).

[2]  S. Kocsis, B. Braverman, S. Roberts, M.J. Stevens, R.P. Mirin, L.K. Shalm, A.M. Steinberg, Science $\underline{332}$ (2011) 1170. K. Edamatsu, R. Shimizu, T. Itoh, Phys. Rev. Lett. $\underline{89}$ (2002) 213601. See also: R. Colella, A.W. Overhauser, S.A. Werner, Phys. Rev. Lett. $\underline{34}$ (1975) 1472. H. Rauch, S.A. Werner, Neutron Interferometry, Clarendon Press, Oxford (2000).

[3]  P.S. Wesson, Five-Dimensional Physics, World Scientific, Singapore (2006). P.S. Wesson, J. Ponce de Leon, J. Math. Phys. $\underline{33}$ (1992) 3883. P.S. Wesson, Gen. Relativ. Gravit. $\underline{40}$ (2008) 1353. L. Randall, R. Sundrum, Mod. Phys. Lett. A $\underline{13}$ (1998) 2807. N Arkani-Hamed, S. Dimopoulos, G. Dvali, Phys. Lett. B $\underline{429}$ (1998) 263.

[4]  B. Mashhoon, H. Liu, P.S. Wesson, Phys. Lett. B $\underline{331}$ (1994) 305. S.S. Seahra, P.S. Wesson, Class. Quant. Grav. $\underline{20}$ (2003) 1321.

[5]  D. Youm, Mod. Phys. Lett. A $\underline{16}$ (2001) 2371. S.S. Seahra, P.S. Wesson, Gen. Relativ. Gravit. $\underline{33}$ (2001) 1731.

[6]  B. Mashhoon, P.S. Wesson, H. Liu, Gen. Relativ. Gravit. $\underline{30}$ (1998) 555. D. Youm, Phys. Rev. D $\underline{62}$ (2000) 084002. A. Bejancu, C. Calin, H.R. Farran, preprint, U. Kuwait (2012).

[7]  P.S. Wesson, Phys. Lett. B$\underline{701}$(2011) 379. P.S. Wesson, Phys. Lett. B $\underline{706}$ (2011) 1.
13